\documentclass{article}
\usepackage{graphicx}
\begin{document}

\title{Line of magnetic monopoles and an extension of the Aharonov-Bohm effect}

\author{J. Chee and W. Lu\\\\School of Science, Department of Physics, \\Tianjin Polytechnic University,
 Tianjin 300387, China}

 \date{February 10, 2016}
 \maketitle
 \begin{abstract}
 In the Landau problem on the two-dimensional plane, magnetic translation of a quantum wave can be induced by an in-plane electric field.
 The geometric phase accompanying such magnetic translation around a closed path differs from the topological phase of Aharonov and Bohm in two essential aspects: The wave is in direct contact with the magnetic flux and the geometric phase has an opposite sign from the Aharonov-Bohm phase. We show that magnetic translation on the two-dimensional cylinder implemented by the Schr\"{o}dinger time evolution truly leads to the Aharonov-Bohm effect. The magnetic field normal to the cylinder's surface is given by a line of magnetic monopoles which can be simulated in cold atom experiments. We propose an extension of the Aharonov-Bohm experiment which can demonstrate the mutually counteracting effect between the local magnetic translation geometric phase and the topological phase of Aharonov and Bohm.
 \end{abstract}
 \pagebreak

\section{Introduction}
The Dirac monopole \cite{dirac} and the Aharonov-Bohm effect \cite{aharonov} are fundamental subjects in quantum physics. They demonstrate how topological conditions may dominate the behavior of quantum systems interacting with the electromagnetic field. They also have intimate connections with the Berry phase (or geometric phase) \cite{berry} which has wide applications in many different areas \cite{shapere} in physics. While the Aharonov-Bohm effect has been verified experimentally, synthetic magnetic fields such as the field of a Dirac monopole or a line of monopoles are explored only recently in experiments involving Bose-Einstein condensates \cite{ray,conduit}. This has opened up new possibilities for quantum physics research.

A unique situation where a line of magnetic monopoles plays an essential role is a charged particle with charge $q$ on a two-dimensional cylinder subjected to a magnetic field $B$ normal to the cylinder's surface, which we refer to as the Landau problem on the cylinder \cite{date}. The magnetic field can be seen as produced by a line of monopoles of uniform density located at the central axis of the cylinder. In order to completely characterize the problem, one needs to specify the vector potential that enters the Hamiltonian. For the same $B$, the vector potential can be classified into gauge equivalence classes according to the magnetic flux $\phi$ that threads the cylinder modulo $hc/q$. (This point was discussed by Wu and Yang \cite{wu-yang} in connection with the Aharonov-Bohm effect. Although in their discussions the line of monopoles is absent, their arguments carry over to the present situation without essential change.) So one needs both $B$ and $\phi$ modulo $hc/q$ to specify a Hamiltonian that belongs to a certain gauge equivalence class.

Laughlin applied the Landau problem on a ribbon, i.e., a finite part of the cylinder in his treatment of the integer quantum Hall effect \cite{laughlin}. Connected with Laughlin's analysis is the fact that the Landau problem on the cylinder has reduced quantum mechanical symmetry so that translation symmetry in the longitudinal direction is discrete. Laughlin's derivation of the integer quantum Hall effect and the reduction of symmetry are both related to the role played by $hc/q$, a quantity that can be reminiscent of the Aharonov-Bohm effect. However, unlike the Aharonov-Bohm effect, they have no bearing on the value $\phi$ modulo $hc/q$. This is ultimately due to the fact that the quantum Hall conductivity and the step size of translation symmetry in the longitudinal direction are changed if $hc/q$ has a different value but neither is affected if one starts out with a different $\phi$. We shall come back to this point later.

Our present work has its origin in the Landau problem on the two-dimensional plane. Imagine a quantum state, say a ground state $\Psi_0$, that is a Gaussian in its probability distribution. Now consider magnetic translation of the wave around a loop on the plane. (The loop can be taken as the trajectory of the center of $\Psi_0$ where the probability density is the greatest. Any point on the wave gives the same trajectory up to translation.) The magnetic translation operator is the usual translation operator times a gauge transformation so that it commutes with the Landau Hamiltonian \cite{brown,zak1,zak2}. It has the same effect as the usual translation operator in shifting the wave's probability distribution. But magnetic translation in two different directions do not commute, and when the wave is brought back to its initial position, it acquires a phase factor $\exp(-iq\phi_B/\hbar c)$, where $\phi_B$ is the magnetic flux $\phi_B$ through the loop.

The above is a kinematical argument based on the magnetic translation symmetry of the Landau Hamiltonian. However, such a magnetic translation around a loop can be implemented by the Schr\"{o}dinger time evolution, if a uniform in-plane electric field is applied that changes direction with time \cite{chee}. The Berry phase factor accompanying such an evolution is $\exp(-iq\phi_B/\hbar c)$. So now the phase factor is not just kinematical but rather has physical consequences if an interference experiment is performed. Observe that there is a difference between the phase factor here and the Aharonov-Bohm phase factor $\exp(iq\phi/\hbar c)$, which has also been derived from a Berry phase approach \cite{berry} in the original context where there is no magnetic field outside a magnetic flux line. One can imagine driving a wave packet around a loop that encircles both a uniform magnetic field and a flux line whose magnetic field points in the same direction. Then $\phi$ and $\phi_B$ have the same sign but the final geometric phase is $\frac{q}{\hbar c}(\phi-\phi_B)$ rather than $\frac{q}{\hbar c}(\phi+\phi_B)$. From a theoretical point of view, the complexity of the circumstance is that the flux line is mixed with the uniform magnetic field and breaks the magnetic translation symmetry of the Landau problem. This may cause complications to a rigorous description of the process if a direction-changing electric field is added.

In addition to the sign difference in $\frac{q\phi}{\hbar c}$ and $-\frac{q\phi_B}{\hbar c}$, $\phi_B$ is different from $\phi$ in that it is not located in a separate region whose contact with the wave can approach zero. On the contrary, at each instant during the evolution, the wave is in direct contact with the uniform magnetic field inside the loop and in this sense it is different from the effect originally proposed by Aharonov and Bohm whose rigorous experimental verification \cite{tonomura} makes a point in completely shielding $\phi$ from the quantum wave. 

Given the above analysis, we wonder what happens in the Landau problem on the two-dimensional cylinder. Since translation symmetry in the longitudinal direction is now discrete, magnetic translation of a wave around a loop can only be that around a cross-sectional circle of the cylinder, which encloses the magnetic flux $\phi$ that does not come in contact with the wave at all. We point out that magnetic translation around such a loop gives exactly $\exp(iq\phi/\hbar c)$ instead of $\exp(-iq\phi/\hbar c)$. Furthermore, we show that such a magnetic translation is implemented by the Schr\"{o}dinger equation on the cylinder so that the Berry phase factor is $\exp(iq\phi/\hbar c)$. So the magnetic translation Berry phase on the cylinder \emph{is} the Aharonov-Bohm phase, both in mathematical form and exact physical content. The simple change in topological conditions has lead to the true connection between magnetic translation and the effect in the original context of Aharonov and Bohm.

The replacement of the uniform magnetic field with the field of a line of monopoles also facilitates a rigorous treatment of the physical situation where $\phi_B$ and $\phi$ have an influence on the quantum evolution at the same time. Their contributions to the total geometric phase tend to cancel out each other. We propose such an experiment involving the magnetic field of a line of monopoles that may represent an interest to experimentalists in cold atom physics.

\section{Magnetic translation on the cylinder and the Aharonov-Bohm effect}
Consider the time-dependent Landau problem on the cylinder with the Hamiltonian
\begin{equation}
H(t)=H_\phi(t)-qE_y(t)y=\frac{1}{2m}\big[\Pi^2_x(t)+ \Pi^2_y\big]-qE_y(t)y,
\end{equation}
where $\Pi_x(t)=p_x-\frac{q}{c}A_x(t)$, $\Pi_y=p_y-\frac{q}{c}A_y$. The vector potential ${\bf A}=(A_x, A_y)$ is given by
\begin{equation}
A_x(t)=-By+\phi(t)/l, \ \ \   A_y=0,
\end{equation}
where $\phi(t)$ is the magnetic flux carried in a flux line threading the cylinder. We assume that the $x$ coordinate axis winds around the cylinder in a circle whose circumference is $l$ and the $y$ axis is also on the cylinder and is perpendicular to the $x$ axis.

Our purpose is to study how the electric field ${\bf E}(t)=(E_x(t), E_y(t))$ induces cyclic motion of an initial eigenstate of $H_\phi(0)$. Note that if $\phi(t)$ is to have the meaning of the magnetic flux threading the cylinder, $E_x(t)$ must be represented by $A_x(t)$ (rather than a scalar potential) through $E_x(t)=-c^{-1}\partial A_x/\partial t=-(lc)^{-1}\partial\phi(t)/\partial t$, as required by Faraday's law.

We adopt the single-valuedness condition for the wave function $\Psi$. Since $(x-l, y)$ and $(x, y)$ represent the same point on the cylinder, it follows that
\begin{equation}
\Psi(x-l,y,0)=\Psi(x,y,0).
\end{equation}
If $U(t)$ is the time-evolution operator generated by $H(t)$, then the single-valuedness of $\Psi(x,y,0)$ implies that $\Psi(x,y,t)=U(t)\Psi(x,y,0)$ is also single-valued, i.e., $\Psi(x-l,y,t)=\Psi(x,y,t)$, for all $t$. This follows from the global single-valuedness of $H(t)$ as an operator, albeit defined using local coordinates endowed by the universal covering map from $R^2$ to the cylinder.

In the literature, the general condition $\Psi(x-l,y,0)=e^{i2\pi\lambda}\Psi(x,y,0)$ is sometimes considered \cite{date,wu}. Then the wave function is not single-valued if $\lambda$ is not an integer. This amounts to a redistribution of the same $\phi$ between the Hamiltonian and the wave function such that $\lambda$ carries some or all of $\phi$. However, the two formulations should give the same physics when the single-valued formulation is applicable.

Considered as a Hamiltonian on a two-dimensional plane $R^2$, $H_\phi(0)$ has magnetic translation symmetry generated by
\begin{equation}
P_x=\Pi_x(0)-\frac{qB}{c}y=p_x-\frac{q\phi}{lc}, \ \ \ P_y=\Pi_y(0)+\frac{qB}{c}x=p_y+\frac{qB}{c}x,
\end{equation}
where $\phi\equiv\phi(0)$. For the Landau problem on $R^2$, $\phi$ is just a choice of gauge that has no physical meaning. Now in the case of the cylinder, the operator
\begin{equation}
M(0, R_y)=\exp(-\frac{i}{\hbar}P_yR_y)=\exp(-\frac{i}{\hbar}p_yR_y)\exp(-\frac{iqBR_yx}{\hbar c}),
\end{equation}
though commuting with $H_\phi(0)$, may not be a single-valued operator. This is because $\exp(-\frac{i}{\hbar}p_yR_y)$ is single-valued but $\exp(-iqBR_yx/(\hbar c))$ is single-valued only when $BlR_y=khc/q$, where $k$ is an integer. In this sense, $H_\phi(0)$ has only discrete magnetic translation symmetry in the $y$ direction if $B\neq 0$. On the other hand, the operator
\begin{equation}
M(R_x, 0)=\exp(-\frac{i}{\hbar}P_xR_x)=\exp(\frac{iqR_x\phi}{\hbar lc})\exp(-\frac{i}{\hbar}p_xR_x)
\end{equation}
commutes with $H_\phi(0)$ and is always well-defined. Thus, continuous magnetic translation symmetry in the $x$ direction is retained. For $R_x=l$, it follows that
\begin{equation}
M(l, 0)\Psi(x, y, 0)=\exp(i\frac{q\phi}{\hbar c})\Psi(x-l, y, 0)=\exp(i\frac{q\phi}{\hbar c})\Psi(x, y, 0).
\end{equation}
Therefore, $M(l, 0)$ produces the phase factor $\exp(i\frac{q\phi}{\hbar c})$ when acting on any wave function defined on the cylinder. This situation is uniquely determined by the magnetic flux $\phi$ and the condition $\Psi(x-l, y, 0)=\Psi(x, y, 0)$, which characterize the topology of the Landau problem on the cylinder, and is independent of the magnetic field $B$ that interacts locally with the quantum wave. In this sense the phase factor is topological. We shall now prove that the phase factor $\exp(i\frac{q\phi}{\hbar c})$ is the consequence of the Schr\"{o}dinger time evolution and is observable. In this sense, it is identical with the Aharonov-Bohm phase factor.

We start from the general case of a time-dependent $\phi(t)$. Let ${\bf R}(t)=(R_x(t), R_y(t))$ be the displacement vector associated with the drift velocity due to the electric field, i.e.,
\begin{equation}
R_x(t)=\frac{c}{B}\int_0^tE_y(\tau)d\tau, \ \ \ R_y(t)=-\frac{c}{B}\int_0^tE_x(\tau)d\tau.
\end{equation}
Combined with $E_x(t)=-(lc)^{-1}\partial\phi(t)/\partial t$, this leads to $\phi(t)=\phi+lBR_y(t)$. Let
\begin{equation}
U(t)=g(t)U'(t), \ \ \ g(t)=\exp\big(\frac{iqBR_y(t)}{\hbar c}x\big).
\end{equation}
Substituting $U(t)$ into the Schr\"{o}dinger equation $i\hbar\dot{U}=HU$, we obtain
\begin{equation}
i\hbar\dot{U}'(t)=H'(t)U'(t), \ \ \ H'(t)=H_\phi(0)-qE_x(t)x-qE_y(t)y.
\end{equation}
The operators $g(t)$ and $U'(y)$ are not single-valued on the cylinder if $BlR_y(t)\neq khc/q$, although their product always is. In analyzing $U(t)$, it is useful to mathematically identify the problem on the cylinder with the corresponding problem on $R^2$ plus the constraint $\Psi(x,y,0)=\Psi(x-l,y,0)$. Viewed as such, both $g(t)$ and $U'(t)$ are legitimate operators. On $R^2$, although $U'(t)$ may produce a quasi-periodic function when acting on the periodic function $\Psi(x,y,0)$ (with period $l$), the factor $g(t)$ necessarily makes it periodic again so that $U(t)$ as a whole always satisfies $U(t)\Psi(x,y,0)=U(t)\Psi(x-l,y,0)$, thus making $\Psi(x,y,t)$ always periodic on $R^2$, or equivalently, single-valued on the cylinder.

The electric field terms in $H'(t)$ can be written in another form
\begin{equation}
-qE_x(t)x-qE_y(t)y={\dot R}_y(t)(P_y-\Pi_y(0))+{\dot R}_x(t)(P_x-\Pi_x(0)),
\end{equation}
Since $H_\phi(0)$ commutes with $P_x$ and $P_y$, we can extract the time-evolution generated by ${\dot R}_x(t)P_x+{\dot R}_y(t)P_y$ and by $H_\phi(0)$ successively, i.e., we let
\begin{equation}
U'(t)=M(C_{{\bf R}(t)})D(t)U_\epsilon(t),
\end{equation}
where $D(t)=e^{-\frac{i}{\hbar}H_\phi(0)t}$ and where
\begin{equation}
M(C_{{\bf R}(t)})=\mathrm{T}\exp[-\frac{i}{\hbar}\int_0^t({\dot R}_x(\tau)P_x+{\dot R}_y(\tau)P_y)d\tau].
\end{equation}
$\mathrm{T}\exp$ represents the time-ordered exponential and as we shall point out, it is determined by the path $C_{{\bf R}(t)}$ traversed by ${\bf R}(t)$ during the time interval $[0, t]$. Substituting $U'(t)$ into $i\hbar\dot{U}'(t)=H'(t)U'(t)$, we get
\begin{equation}
i\hbar\dot{U}_\epsilon(t)=H_\epsilon(t)U_\epsilon(t),
\end{equation}
where
\begin{equation}
H_\epsilon(t)=D^{-1}(t)[-{\dot R}_y(t)\Pi_y(0)-{\dot R}_x(t)\Pi_x(0)]D(t).
\end{equation}
Let $R(t)=(R_x(t)+iR_y(t))/\sqrt{2}$ and $\Pi(0)=(\Pi_x(0)+i\Pi_y(0))/\sqrt{2}$. Then we have $D^{-1}(t)\Pi(0)D(t)=\exp(-i\omega t)\Pi(0)$ where $\omega=qB/(mc)$, which is the solution to the Heisenberg equation $i\hbar{\dot\Pi}=[\Pi, H_\phi(0)]$ for $\Pi(t)$. So we get
\begin{equation}
H_\epsilon(t)=-{\dot R}^*(t)e^{-i\omega t}\Pi(0)-{\dot R}(t)e^{i\omega t}\Pi^\dagger(0).
\end{equation}
Let us assume that $R(t)$ depends on $t$ through a slowness parameter $\epsilon$ as $R(\epsilon t)$. In the limit that $\epsilon/\omega \rightarrow 0$, the quantum adiabatic theorem implies that the oscillating $\exp(-i\omega t)$ and $\exp(i\omega t)$ have an averaging effect so that a small ${\dot R}(\epsilon t)$ does not accumulate over a period during which $R(\epsilon t)$ makes a finite change \cite{messiah}. So if $R(\epsilon t)$ changes adiabatically, then $H_\epsilon(t)\rightarrow 0$, which implies that $U_\epsilon(t)$ approaches the identity operator $I$.

In the special case where $\phi(t)=\phi(0)\equiv\phi$, or equivalently, $E_x(t)=0$, we have $R_y(t)=0$. This implies that $g(t)=I$ and $M(C_{{\bf R}(t)})=M(R_x(t), 0)$. If in addition $R_x(t)$ changes adiabatically, then $U_\epsilon(t)\rightarrow I$. So we have $U(t)=M(R_x(t), 0)D(t)$. If $\Psi_n(x,y,0)$ is an eigenstate of $H_\phi(0)$ with eigenvalue $E_n=\hbar\omega(n+1/2)$, then its adiabatic time evolution is given by
\begin{equation}
U(t)\Psi_n(x,y,0)=\exp(-\frac{i}{\hbar}E_nt)\exp(\frac{iqR_x(t)\phi}{\hbar lc})\Psi_n(x-R_x(t),y,0).
\end{equation}
From this we see that the quantum state moves in the positive direction of the $x$ axis if $R_x(t)$ increases with $t$. In particular, when $R_x(T)=l$ at $t=T$, it returns to the initial state, by virtue of the single-valuedness of $\Psi_n(x,y,0)$, acquiring the Berry phase factor $\exp(i\frac{q\phi}{\hbar c})$ in addition to the dynamical phase factor $\exp(-\frac{i}{\hbar}E_nT)$.

The result implies that magnetic translation on the two-dimensional cylinder truly leads to the Aharonov-Bohm effect. The simple change in topological conditions has dominated a change of physics from the planar case \cite{chee} to the cylinder case. One can imagine splitting a quantum wave into two components that stay in two close but separate cylinders made of thin tube-like materials both subjected to the same radial magnetic field $B$. Cyclic motion of one component can be achieved by applying an electric field $E_y$ on one tube only. At the end of the cycle, interference of the two components can detect the Aharonov-Bohm phase factor in this unique circumstance.  Note that continuous magnetic translation on the cylinder is induced by the electric field in the $y$ direction only. This can be an advantage since it might be experimentally feasible to use an effective electric field such as gravity that does not change direction with time to study properties of the exotic magnetic field produced by a line of Dirac monopoles that is now of relevance in cold atom experiments.

\section{$\exp(i\frac{q\phi}{\hbar c})$ versus $\exp(-i\frac{q\phi_B}{\hbar c})$}

In a general adiabatic time evolution governed by $H(t)$, the flux $\phi(t)$ is time-dependent, which leads to a non-vanishing $E_x(t)$, or equivalently, a time-dependent $R_y(t)$. As a result, the quantum wave can move in the $y$ direction as well as in the $x$ direction. Such a general adiabatic time evolution leads to a natural extension of the Aharonov-Bohm effect which brought together the topological phase factor $\exp(i\frac{q\phi}{\hbar c})$ and the phase factor $\exp(-i\frac{q\phi_B}{\hbar c})$ that arises from local interaction of the quantum wave with the magnetic field $B$.

For the general adiabatic evolution, we have $U(t)=g(t)M(C_{{\bf R}(t)})D(t)$. While $g(t)M(C_{{\bf R}(t)})$ is well-defined on the cylinder, $M(C_{{\bf R}(t)})$ as an intermediate step should, in general, be seen as defined on $R^2$. To calculate the time-ordered exponential in $M(C_{{\bf R}(t)})$, consider two successive magnetic translations on $R^2$ corresponding to the vectors ${\bf R}_1$ and ${\bf R}_2$. The formula $e^Ae^B=e^{A+B}e^{\frac{1}{2}[A,B]}$, which applies when $A$ and $B$ both commute with $[A, B]$, gives
\begin{eqnarray}
M({\bf R}_2)M({\bf R}_1)&=&e^{-\frac{i}{\hbar}{\bf P}\cdot({\bf R}_1+{\bf R}_2)}e^{\frac{1}{2}[-\frac{i}{\hbar}{\bf P}\cdot{\bf R}_2, -\frac{i}{\hbar}{\bf P}\cdot{\bf R}_1]},\nonumber\\&=&M({\bf R}_1+{\bf R}_2)e^{-i\frac{qB}{\hbar c}\frac{1}{2}({\bf R}_1\times{\bf R}_2)\cdot{\bf n}},
\end{eqnarray}
where $[P_x, P_y]=-i\hbar qB/c$ and ${\bf e}_x\times{\bf e}_y=\bf n$ are used. $M(C_{{\bf R}(t)})$ can be calculated by dividing $C_{{\bf R}(t)}$ into small segments. By repeated application of the above formula and taking the zero limit for each segment, we have
\begin{eqnarray}
M(C_{{\bf R}(t)})=e^{-i\frac{q\phi_B}{\hbar c}}M({\bf R}(t)), \ \ \ \phi_B=BS=B\big[\frac{1}{2}\int_0^t{\bf R}(\tau)\times d{\bf R}(\tau)\cdot{\bf n}\big].
\end{eqnarray}
Clearly, $S$ is the oriented area swept by ${\bf R}(t)$ during the time interval $[0, t]$. It is equal to the area enclosed by $C_{{\bf R}(t)}$ and the vector $-{\bf R}(t)$ which points from the point ${\bf R}(t)$ to ${\bf R}(0)={\bf 0}$. Clearly, any path $C_{{\bf R}(t)}$ can be decomposed into a closed path $C_{{\bf R}(t)}+(-{\bf R}(t))$ and the line segment ${\bf R}(t)$ pointing directly from the point ${\bf R}(0)={\bf 0}$ to the point ${\bf R}(t)$. In the case ${\bf R}(t)=(l, 0)$, the line segment on $R^2$ corresponds to a closed path on the cylinder so we have one closed path $C_{{\bf R}(t)}$ decomposed into a contractible loop $C_{{\bf R}(t)}+(-{\bf R}(t))$ and a circle $(l, 0)$ that is not contractible.

In order that an initial eigenstate $\Psi_n(x,y,0)$ returns to itself at $t=T$, $R_y(T)$ must return to $R_y(0)=0$ which implies $g(T)=1$. However, $R_x(T)$ does not have to return to $0$. In particular, for ${\bf R}(T)=(l, 0)$, we have
\begin{eqnarray}
M(C_{{\bf R}(T)})D(T)\Psi_n(x,y,0)&=&e^{-i\frac{q\phi_B}{\hbar c}}M(l, 0)D(T)\Psi_n(x,y,0)\nonumber\\&=&e^{-\frac{i}{\hbar}E_nT}e^{i\frac{q}{\hbar c}(\phi-\phi_B)}\Psi_n(x,y,0).
\end{eqnarray}
So in this case we have a cyclic evolution and the Berry phase is $\frac{q}{\hbar c}(\phi-\phi_B)$. The Berry phase would be the same if the wave is driven on the cylinder around $C_{{\bf R}(T)}+(-{\bf R}(T))$ followed by ${\bf R}(T)$. The former encloses only the flux due to $B$ and contributes $-\frac{q}{\hbar c}\phi_B$ while the latter has been identified in the previous section as leading to the Aharonov-Bohm phase $\frac{q}{\hbar c}\phi$.

In the above conclusions, we find an extension of the Aharonov-Bohm interference experiment that would verify the different contributions of $\phi$ and $\phi_B$ to the total geometric phase. 
\begin{figure}[h]
  \centering
  \includegraphics{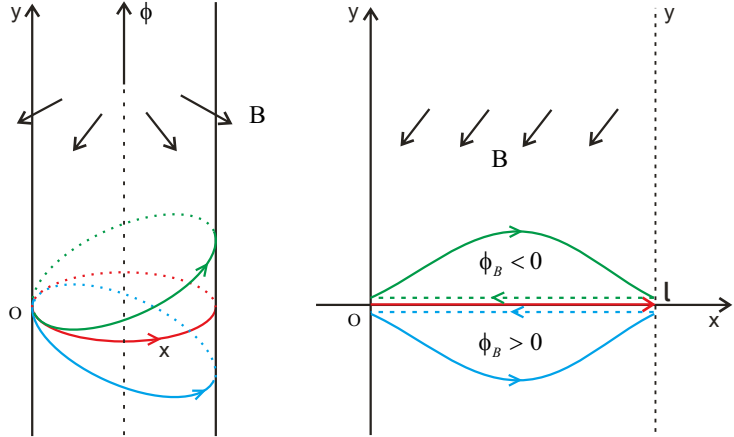}
  \caption{A wave packet can be transported along either the green or the blue loop. Each loop is the sum of a contractible loop (drawn on $R^2$ in a different scale) and the red circle. The total flux through the blue loop is $\phi+\phi_B>0$. Yet the geometric phase is 0 if $\phi=\phi_B$. Likewise, the total flux through the green loop is 0 when the geometric phase is $2q\phi/(\hbar c)$. An interference experiment can demonstrate the counteracting effect between the ordinary flux due to $B$ and the Aharonov-Bohm flux that is not in contact with the wave.}\label{Fig}
\end{figure}
Consider two different loops on the cylinder marked blue and green in Figure 1. The magnetic field due to the monopole line points outward and that of the flux line points upward. Each loop, when decomposed into the two loops discussed above, contains the common red circle that encloses $\phi$. But the other two contractible loops (drawn on the covering space) have different orientations. As a result, they enclose two different $\phi_B$'s with opposite signs. This leads to different total geometric phases. For instance, although the blue loop encloses more total flux, the geometric phase they contribute cancel each other out when $\phi=\phi_b$. This can potentially be verified in experiment. This paragraph has summarized the main physical results of our paper.

In order to achieve the general adiabatic evolution, one needs a time-dependent $\phi(t)$ in addition to an electric field in the $y$ direction and a line of magnetic monopoles. These seem to be feasible with developments of experimental technology in cold atom physics.

\section{Comparison with Laughlin's thought experiment, the spin experiment and discussions}
Date and Divakaran \cite{date} have studied the time-independent problem on the cylinder with $E_y\equiv 0$ from a mathematical perspective. Laughlin \cite{laughlin} on the other hand considered the case where $E_y(t)\equiv -E_0\neq 0$. He imagined a ``charge pump" where a slow change of $\phi$ causes an initial eigenstate of $H(0)$ to move in the $y$ direction, which he used as a theoretical tool to derive the integer quantum Hall effect. The electric field $-E_0$ in Laughlin's case establishes a net sum $I$ of individual currents carried by eigenstates $e^{i\kappa x}\phi_n(y-y_0)$ of $H(0)$, where $\kappa$ is the discrete wave vector, $\phi_n$ is the $n$th eigenstate of the harmonic oscillator, and $y_0$ is linear in $\phi$, $\kappa$ and $E_0$. Under adiabatic variation of $\phi$, only $y_0$ is changed. These eigenstates stand still in the $x$ direction. Even when one considers a superposition of different $\kappa$'s with the same $n$, this is still true because the phase differences among these eigenstates cannot change. So within each $n$, quantum interference caused by nontrivial cyclic motion of a wave around the cylinder cannot happen. There is the possibility of using superposition of eigenstates with different $n$'s. However, this is beyond the main theme of the quantum Hall effect and as far as we know, has not been considered in the literature. Let us also note that when $\phi$ changes by $hc/q$, Laughlin's charge pump moves the system along $y$ by the distance $hc/(qBl)$ that is the step size of discrete translation symmetry, corresponding to just enough area each quantum state must occupy. The process is insensitive to what value of $\phi$ modulo $hc/q$ one starts out and ends up with.

Here we allow $E_y(t)$ to be time dependent. In this sense it is more general. Yet our main difference compared with Laughlin's charge pump is that we ask what happens if we prepare an eigenstate of $H_\phi(0)$ rather than $H(0)$, and then turn on the electric field so that the state evolves under $H(t)$. An eigenstate of $H_\phi(0)$ can easily be chosen as a wave concentrated rather than spread uniformly around in the $x$ direction. Let $\Psi_n(x,y,0)$ be a quantum wave in a magnetic field which is an eigenstate of $H_\phi(0)$ considered on the two-dimensional plane. Increasing the magnetic field can reduce the size of the wave where the probability density is significantly different from zero. Then
$$C\sum^{\infty}_{k=-\infty}\exp(-ip_xkl)\Psi_n(x,y,0),$$
where $C$ is a normalization constant, is a periodic function on $R^2$ with period $l$ which is readily identifiable as an eigenstate of $H_\phi(0)$ on the cylinder with the same eigenvalue. Note that $\exp(-ip_xkl)$ commutes with $H_\phi(0)$.

The quantum adiabatic theorem is readily connected to the problem we have considered.
Through the time-dependent gauge transformation,
\begin{equation}
\Psi(x,y,t)\rightarrow \Psi'(x,y,t)=\exp(\frac{iqBR_x(t)y}{\hbar c})\Psi(x,y,t),
\end{equation}
$H(t)$ is correspondingly transformed into
\begin{equation}
H'(t)=\frac{1}{2m}\big[(p_x-\frac{q}{c}A_x(t))^2+ (p_y-\frac{q}{c}(-BR_x(t)))^2\big].
\end{equation}
Note that $H'(0)=H_\phi(0)$. Since $\phi(t)=\phi(0)+lBR_y(t)$, $H'(t)$ depends on time through $(R_x(t), R_y(t))$. In this gauge the problem is how $H'(t)$ evolves its own initial eigenstate. The definition of the dynamical phase factor $\exp(-\frac{i}{\hbar}E_nT)$ is natural seen in this gauge since the energy eigenvalues of $H'(t)$ do not change with time.

In this paper, we have focused on the Landau problem assuming the particle has spin 0. In a recent experiment, a mesoscopic ring of layered semiconductor containing a two-dimensional electron gas is used to study simultaneously the Aharonov-Bohm, the Aharonov-Casher and the spin Berry phase effects \cite{nagasawa}. In this experiment, the Rashba spin-orbit interaction couples the electron spin to an effective radial magnetic field perpendicular to the electron momentum along the ring. The ring, however, lies in the plane perpendicular to $\phi$, and thus does not include the effect we have studied in this paper. Adding a nonzero spin to our investigation would lead to the same additional effects as studied in this experiment.

In general, a cyclic adiabatic evolution requires $R_y(T)=0$, but $R_x(T)$ can return to any integer multiples of $l$. This is similar to Zak's study \cite{zak} of nontrivial Berry phase associated with driving a Bloch state around a loop in a Brillouin zone where the change of the Bloch quasimomentun is integer multiples of $2\pi/a$, where $a$ is the lattice constant.

Similar to $M(C_{{\bf R}(T)})$, the operator $U_\epsilon(t)$ can also be calculated on $R^2$ \cite{chee} for an arbitrary change of ${\bf R}(t)$ that is not necessarily adiabatic. So the time evolution is explicitly known for any given initial state that may or may not be an eigenstate of $H_\phi(0)$. This may lead to a rigorous identification of the non-adiabatic geometric phase proposed by Aharonov and Anandan \cite{aa} and the Aharonov-Bohm phase.

Recent advances in cold atom physics have also made possible the study of synthetic nonabelian gauge fields through experiments. Of particular interest is the nonabelian version of the Landau problem \cite{estienne} which exhibits rich physical phenomena. It might be of significance to generalize the question we have studied here to the nonabelian case in connection with the nonabelian generalization of the Aharonov-Bohm effect \cite{wu-yang}.

In conclusion, by replacing the two-dimensional plane with the two dimensional cylinder, we find the true connection between magnetic translation and the topological effect originally proposed by Aharonov and Bohm. The line of magnetic monopoles required in the cylinder case is physically realizable in cold atom physics. An extension of the Aharonov-Bohm experiment can verify the combined effects of the local magnetic translation geometric phase and the Aharonov-Bohm phase. As far as we know, no experiment has been done to test such combined effects on a quantum wave when both $\phi$ and $\phi_B$ are present. We note that in the literature, the term Aharonov-Bohm effect is sometimes used to refer to the geometric phase due to the local interaction with a magnetic field, which we in this paper have referred to as the magnetic translation geometric phase. This effect is the focus of recent investigations involving neutral cold atoms interacting with laser beams \cite{dalibard}.

\end{document}